\documentclass[
aip,
rsi,
amsmath,amssymb,
reprint,
]{revtex4-1}

\usepackage{graphicx}
\usepackage{dcolumn}
\usepackage{bm}
\usepackage{float}
\usepackage[caption = false]{subfig}
\usepackage{natmove}

\newcommand{\kB}{k_{\text{B}}}
\newcommand{\tobs}{t_{\text{obs}}}
\newcommand{\traj}{\Psi(\tobs)}
\newcommand{\op}[1]{\hat{#1}}
\newcommand{\ket}[1]{| #1 \rangle}
\newcommand{\bra}[1]{\langle #1 |}
\newcommand{\braket}[1]{\langle #1 \rangle}
\newcommand{\thavg}[1]{\langle #1 \rangle_{\text{eq}}}
\newcommand{\qexpsym}[2]{ \langle #1 \rangle_{#2}}
\newcommand{\qexp}[2]{\bra{\psi_{#2}} \op{#1} \ket{\psi_{#2}}}
\newcommand{\peavg}[1]{\left <{#1} \right >}
\newcommand{\cpeavg}[2]{\left <{#1} \right >_{#2}}

\begin{document}

\preprint{AIP/123-QED}

\title{Studying rare nonadiabatic dynamics with transition path sampling quantum jump trajectories}

\author{Addison J. Schile}
\affiliation{Department of Chemistry, University of California, Berkeley}
\affiliation{Lawrence Berkeley National Laboratory, University of California, Berkeley}
\author{David T. Limmer}
 \email{dlimmer@berkeley.edu.}
\affiliation{Department of Chemistry, University of California, Berkeley}
\affiliation{Kavli Energy NanoSciences Institute, University of California, Berkeley}
\affiliation{Lawrence Berkeley National Laboratory, University of California, Berkeley}

\date{\today}

\begin{abstract}

We present a method to study rare nonadiabatic dynamics in open quantum systems using transition path sampling and quantum jump trajectories.
As with applications of transition path sampling to classical dynamics, the method does not rely on prior knowledge of transition states or reactive pathways, and thus can provide mechanistic insight into ultrafast relaxation processes in addition to their associated rates. 
In particular, we formulate a quantum path ensemble using the stochastic realizations of an unravelled quantum master equation, which results in trajectories that can be conditioned on starting and ending in particular quantum states. Because the dynamics rigorously obeys detailed balance, rate constants can be evaluated from reversible work calculations in this conditioned ensemble, allowing for branching ratios and yields to be computed in an unbiased manner. 
We illustrate the utility of this method with three examples: energy transfer in a donor-bridge-acceptor model, and models of photo-induced proton-coupled electron transfer and thermally activated electron transfer.
These examples demonstrate the efficacy of path ensemble methods and pave the way for their use in studying of complex reactive quantum dynamics.

\end{abstract}

\pacs{Valid PACS appear here}

\maketitle

\section{\label{sec:intro}Introduction}

Understanding the dynamics of quantum systems in condensed phases is an active area of research across physics and chemistry~\cite{mukamel2000multidimensional,fayer2009dynamics,ishizaki2012quantum,jang2018delocalized}. 
Advances in time-resolved spectroscopies, such as pump-probe transient absorption and coherent two-dimensional spectroscopy, have made it possible to measure dynamics on ultrafast timescales,~\cite{krvcmavr2014signatures,duan2016quantum,oliver2014correlating,monahan2017room} but require sophisticated simulation methodologies to help interpret and unravel the microscopic motions probed~\cite{ikeda2017probing,fetherolf2017linear}. 
In this paper, we demonstrate how the Transition Path Sampling (TPS)\cite{dellago2002transition,bolhuis2002transition} framework can be used effectively for studying the dynamics of nonadiabatic quantum systems. 
We do this by taking advantage of a stochastic trajectory representation of a detailed-balance-preserving quantum master equation, which allows for the generation of a trajectory ensemble whose statistics and correlations can be studied.
We show how the use of path ensembles can elucidate dynamical mechanisms directly using a generalization of committor analysis~\cite{dellago2002transition,bolhuis2002transition,peters2016reaction} for coherent dynamics. 
Additionally, we show how TPS can be used to compute rate constants for rare dynamical events without assuming a specific mechanism or postulating a relevant reaction coordinate using path ensemble free energies~\cite{dellago2002transition,bolhuis2002transition}. 
While the current framework is restricted to quantum jump dynamics, the perspective is general and the tools are generalizable to other open quantum dynamics~\cite{montoya2015extending,walters2016iterative}.

Nonadiabatic open quantum systems display a wide variety of chemical physics, from excitonic behavior in chromophoric systems\cite{ishizaki2012quantum,jang2018delocalized} to conical intersections\cite{domcke2004conical,hahn2002ultrafast} and span a number of time, energy, and length scales\cite{tully2012perspective}. This vast range of scales makes developing computational techniques for studying nonadiabatic dynamics  difficult. The break down of the Born-Oppenheimer approximation necessitates that the dynamical evolution of the system is based on Schrodinger's equation, while the surrounding environment necessary to accurately describe dissipation makes its straightforward application intractable due to the exponential scaling with system size. Thus, most numerical techniques are developed to treat a few degrees of freedom quantum mechanically, resolving discrete electronic states or wavepackets, while the other degrees of freedom are treated with path integrals\cite{topaler1996path},  or approximately semi-classically\cite{miller1970classical,ben1998nonadiabatic}, with mixed quantum-classical dynamics\cite{kapral1999mixed,mclachlan1964variational,tully1990molecular}, or with reduced density matrix equations\cite{micha2012density,kelly2013efficient,montoya2015extending}.

Independent of the computational technique, the prevailing perspective for studying such nonadiabatic open quantum dynamics relies on computing and analyzing the average dynamics that a few tagged degrees of freedom undergo when the rest of the system has been integrated out. This is sometimes done implicitly, by focusing on average populations even though the surroundings are represented molecularly, as is done with semiclassical methods~\cite{lee2016semiclassical,tully1990molecular,topaler1996path,bonella2005linearized}. Often, however, this is done explicitly, as in methods that construct an equation of motion for the average behavior of the system directly, as in quantum master equation approaches~\cite{hahn2002ultrafast,ishizaki2009theoretical,berkelbach2013microscopic,tempelaar2018vibronic,berkelbach2012reduced,kelly2013efficient}. While this dimensionality reduction can be illuminating, it does result in a loss of information, as the fluctuations about the average dynamical behavior can encode important correlations. For example, understanding the mechanism of a rare dynamical event with information on just the average trajectory of the system is difficult. Most often a mechanism is inferred by varying a parameter of the system and noting the subsequent change in the rate. Instead of noting the response to a parameter, this same information exists in principle in the ensemble of trajectories, or dynamical fluctuations of the system, at a fixed value of a parameter. In static systems this is just a statement of the fluctuation-dissipation relation, such statements can be extended to codify the relation between fluctuations and response in dynamical systems far from equilibrium~\cite{touchette2009large,seifert2010fluctuation}.
Indeed in classical systems, trajectory ensemble techniques have resulted in methods like Transition Path Sampling\cite{dellago2002transition,bolhuis2002transition} to sample rare dynamical events and generalizations of reaction coordinates and transitions state to complex systems~\cite{dellago2002transition,bolhuis2002transition,peters2016reaction,hummer2004transition}. This has enabled the study 
of mechanisms of rare events in a wide variety of systems and settings~\cite{geissler1999kinetic,kattirtzi2017microscopic,best2005reaction,geissler2001autoionization}. 

The application of trajectory ensembles to quantum dynamics, however, has not been as successful as in classical dynamics. 
Central to this failure is the difficulty in generating meaningful trajectories for open quantum systems. 
For many trajectory-based methods, the dynamics are reliable only for very short times due either to approximations that fail to accurately represent the back-reaction of the bath onto the system and consequently violate detailed balance~\cite{kapral2015quantum} or because of the dynamical sign problem and exponential complexity of exact system-bath dynamics.
Alternatively, path integral methods such as recent extensions to Ring Polymer Molecular Dynamics\cite{habershon2013ring} that incorporate non-adiabatic effects\cite{menzeleev2014kinetically,shushkov2012ring,ananth2013mapping,richardson2013communication,tao2018path} can recover the correct equilibrium statistics, and could be used to generate quantum trajectory ensembles in cases where they are also faithful to the quantum dynamics.  
Efforts to use practical methods such as surface hopping with trajectory ensembles have been proposed~\cite{sherman2016nonadiabatic} though their reliability is questionable, as the form of the stationary distribution is unknown, making deriving acceptance criteria difficult. Recent work to identify an incompressible phase space structure for the density matrix of an open quantum system in the presence of quenched disorder holds significant promise~\cite{dodin2018approaching}.

In cases where the system and bath are weakly coupled however, the stochastic unraveling method from quantum optics as applied to quantum master equations supplies a means to identify quantum trajectories\cite{breuer1995stochastic,plenio1998quantum}.
In this method, a deterministic density matrix equation is converted to an average over stochastically evolved wavefunctions. Provided a microscopic model of the system bath interaction, the stochastic evolution be can developed.
Such quantum trajectories are observable in simple systems using weak measurements~\cite{murch2013observing}. 
A significant amount of work has been done using quantum jump trajectories in driven systems and under steady-state conditions, which have revealed the potential for dynamical phase transitions~\cite{garrahan2010thermodynamics,ates2012dynamical}, correlated dynamics~\cite{lan2018quantum}, and localization~\cite{levi2016robustness}.
Here we adopt this perspective and develop it with the motivation to study rare reactive events in nonadiabatic and quantum coherent dynamics. As this method is derived from a quantum master equation formalism, its dynamics obey detailed balance, and so its statistical fluctuations encode accurate information on the bath fluctuations that result in rare reactive events. While the bath is not represented in molecular detail, the fluctuations it imposes on the systems dynamics are directly observable. 

The remainder of this paper is outlined in five sections. In the following section, the trajectory formalism is introduced and the formulation of path ensembles and a scheme to sample them with TPS is developed. 
This path ensemble formalism is then applied to three different model systems: first to a three-level chromophoric system to show how path ensembles can be used to sample correlations in trajectories directly (Sec. \ref{sec:cond}), then to a proton-coupled electron transfer model in which the quantum committor distribution is utilized (Sec. \ref{sec:comm}), and finally to a system exhibiting rare barrier crossing to show the efficiency of TPS to compute a rate constant with no mechanistic assumptions (Sec. \ref{sec:rates}). 
Some final conclusions and thoughts for future work are presented in Sec. \ref{sec:conclusions}.

%%%%%%%%%%%%%%%%%%%%%%%%%%%%%%%%%%%%%%%%%%%%%%%%%%%%%%%%%%%%%%%%%%%%%%%%%%%%%
% Theory Section
%%%%%%%%%%%%%%%%%%%%%%%%%%%%%%%%%%%%%%%%%%%%%%%%%%%%%%%%%%%%%%%%%%%%%%%%%%%%%
\section{\label{sec:theory}Quantum Jump Path Ensembles}

In this section we develop a reactive path ensemble formalism for stochastic quantum jump dynamics\cite{breuer2002theory}. Specifically, we consider the reduced dynamics of a subset of degrees of freedom, the system, embedded in an environment with an infinite number of degrees of freedom, the bath, and focus our discussion to instances where those reduced dynamics are Markovian and weakly coupled to the environment. 
For concreteness we will consider Hamiltonians in the full Hilbert space, $\op{H}$, partitioned into three terms,
\begin{equation}
\op{H} = \op{H}_S + \op{H}_B + \op{H}_{SB},
\label{eq:ham_partition}
\end{equation}
where $\op{H}_S$ is the system Hamiltonian, $\op{H}_B$ is the bath Hamiltonian, and $\op{H}_{SB}$ is the system-bath coupling term. Throughout, we will take $\op{H}_{SB}$ as a sum of Kronecker products of linear operators in the system and bath Hilbert spaces,
\begin{equation}
\op{H}_{SB} = \sum_{i} \sum_n c_{n,i} \op{s}_i \otimes \op{B}_{n,i},
\end{equation}
where $\op{s}_i$ is a system operator, and $\op{B}_{n,i}$ the corresponding bath operator. The coefficient $c_{n,i}$ relates the local system-bath coupling strength and in the case where the bath is harmonic, it is convenient to introduce the spectral density,
\begin{equation}
\label{Eq:SpecDens}
\mathcal{J}_i(\omega) = \frac{\pi}{2} \sum_n c_{n,i}^2 \delta (\omega - \omega_n).
\end{equation}
as the weighted sum of the system-bath coupling strengths and density of states at bath frequency $\omega_n$. The spectral density can be inferred from linear absorption measurements~\cite{nitzan2006chemical} or computed from atomistic simulations.~\cite{lee2016modeling}. 

\subsection{Stochastic Wavefunctions from Quantum Jumps}

Provided the Markovian, weak coupling, and secular assumptions, trajectories traced out by the system degrees of freedom consist of periods of coherent evolution punctuated by abrupt changes in the state of the system, reflecting the instantaneous action of the bath. These trajectories represent physical realizations of a piecewise deterministic stochastic process in a projective Hilbert space\cite{breuer1995stochastic} and provide a theoretical description of quantum jump observations in experiments.~\cite{wiseman1993interpretation,goan2001continuous}
The time evolution for a wavefunction in the system Hilbert space over a quantum jump trajectory is given by the stochastic equation of motion,
\begin{align}
\label{Eq:SDE}
d |\psi_t \rangle &= -\frac{i}{\hbar} \op{H}_{\text{eff}} |\psi_t\rangle dt \nonumber \\
&+ \sum_n \left( \frac{\sqrt{\Gamma_n} \op{L}_n}{\langle \psi_t | \Gamma_n \op{L}_n^{\dagger} \op{L}_n |\psi_t \rangle} - 1 \right) |\psi_t \rangle dN_n,
\end{align}
where $\ket{\psi_t}$ is the wavefunction of the system at time $t$ and $\hbar$ is Planck's constant divided by 2$\pi$. The first term in Eq.~\ref{Eq:SDE} represents coherent, deterministic dynamics with the effective Hamiltonian, $\hat{H}_{\text{eff}}$,
\begin{equation}
\hat{H}_{\text{eff}} = \hat{H}_S - \frac{i}{2} \sum_n \Gamma_n \hat{L}_n^{\dagger} \hat{L}_n,
\end{equation}
which adds to the original Hermitian operator, $\op{H}_S$, an anti-Hermitian term due to the coupling with the bath through the operators $\op{L}_n$ and their adjoints, $\hat{L}_n^{\dagger}$. The $\op{L}_n$ operators, include both dissipative and dephasing actions of the bath and $\Gamma_n$ are the associated bare rates of those actions.
The second term in Eq.~\ref{Eq:SDE} is a Poisson jump process reflecting projective actions of the bath with statistics $dN_n = 0,1$ and $dN_n^2 = dN_n$ and rates for each $\op{L}_n$ corresponding to the quantum expectation, $\Gamma_n \bra{\psi_t} \op{L}^{\dagger}_n \op{L}_n \ket{\psi_t}$.

When averaged over a large number of realizations Eq.~\ref{Eq:SDE} returns a master equation describing the probability flow of the Poisson stochastic process, which is of Lindblad form\cite{lindblad1976generators,gorini1976completely},
\begin{align}
\label{eq:lbme}
\partial_t \op{\sigma} (t) &= -\frac{i}{\hbar} [\op{H}_S, \op{\sigma} (t)] \nonumber \\
&+ \sum_n \Gamma_n \left( \op{L}_n \sigma (t) \op{L}_n^{\dagger} - \frac{1}{2} \{ \op{L}_n^{\dagger} \op{L}_n, \op{\sigma} (t) \} \right),
\end{align}
where $\partial_t \op{\sigma}(t)$ is the time derivative of the reduced density matrix, $\op{\sigma}$, and $[\cdot,\cdot]$ is the commutator and $\{\cdot,\cdot\}$ the anti-commutator. Because the system and bath are weakly coupled, each stochastic trajectory is independent and the density matrix is obtainable from the stochastic wavefunctions by $\sigma(t) = \peavg{\ket{\psi_t}\bra{\psi_t}}$ where the brackets denote an average over the Poisson random noise. This master equation is known to form a dynamical semigroup, so that the equation of motion conserves the norm and positivity of the reduced density matrix.\cite{lindblad1976generators,gorini1976completely,breuer2002theory} The semigroup property is vital for a trajectory analysis as it ensures each trajectory has physical meaning and can be experimentally realized.\cite{breuer2004genuine} Stochastic equations of motion have been previously developed for a number of quantum master equations\cite{kleinekathofer2002stochastic,kondov2003stochastic}, however, the representation often gives unphysical trajectories stemming from the underlying master equation's failure to form a dynamical semigroup. Additionally, stochastic unraveling has the algorithmic benefit of reduced scaling in propagating wavefunctions compared to propagating density matrices\cite{pollard1994solution}, which takes the overall scaling in terms of the number of system states $N$ from $\mathcal{O} (N^3)$ to $\mathcal{O} (M N^2)$ where $M$ is the number of trajectories required to converge the density matrix.

The operators, $\op{L}_n$, are identified as Lindblad operators and can be obtained directly from the original system-bath coupling operators\cite{vogt2013stochastic,jeske2013derivation} of a microscopic model provided non-secular terms that couple populations and coherences are negligible\cite{balzer2005modeling}. In this case, it will be most convenient to represent the Lindblad operators in the energy eigenbasis, denoted $\op{L}_{ij}$, and are given by
\begin{equation}
\label{eq:lbops}
\op{L}_{ij} = \op{\mathcal{P}}_{ij} \op{s}
\end{equation}
where $\op{\mathcal{P}}_{ij}$ is an operator that projects out the $ij$ elements of the system-bath coupling operator in the energy eigenbasis, \textit{i.e.}, $\op{\mathcal{P}}_{ij} \op{s} = s_{ij} \ket{\phi_i}\bra{\phi_j}$, where $\ket{\phi_i}$ is the $i$th energy eigenfunction of $\op{H}_S$. The associated rates in the energy eigenbasis, $\Gamma_{ij}$, are given by the Fourier-Laplace transform of the bath correlation function
\begin{equation}
\label{eq:lbrates}
\Gamma_{ij} = \int_0^{\infty} dt \, e^{-i \omega_{ij} t} \thavg{B(t) B(0)},
\end{equation}
where $\thavg{\cdots}$ is a thermal average and $\omega_{ij}=(E_i - E_j)/\hbar$ where $E_{i} \, (E_j)$ is the $i$th ($j$th) eigenvalue of $\op{H}_S$~\cite{nitzan2006chemical}. By the fluctuation-dissipation theorem for quantum time-correlation functions, these rates thus obey detailed balance,
\begin{equation}
\label{eq:detbal}
\frac{\Gamma_{ij}}{\Gamma_{ji}} = e^{\beta \hbar \omega_{ij}},
\end{equation}
where $\beta = 1 / \kB T$ is inverse temperature, $T$, times Boltzmann's constant, $\kB$. This ensures that in the long time limit, the density matrix is given by a Gibbs state, $\hat{\sigma} = \sum_i e^{-\beta E_i} \ket{\phi_i}\bra{\phi_i} $. The Lindblad operators for nonzero frequencies, which are non-diagonal, are associated with population transfer while the zero frequency Lindblad operators, which are diagonal, are the dephasing operators~\cite{breuer2002theory}.

\subsection{Reactive Path Ensembles}

Provided the stochastic equation of motion for the system wavefunction, we can define an ensemble of trajectories parameterized by a trajectory length $\tobs$. This follows closely previous work considering the spacetime thermodynamics of quantum jump processes~\cite{garrahan2010thermodynamics}. We define a sequence of wavefunctions visited over the observation time, $\Psi (\tobs) = \{ \ket{\psi_0} , \ket{\psi_{\Delta t}} , \ldots , \ket{\psi_{\tobs}}  \}$ and the probability of observing that sequence, $P[ \traj ]$, is given by 
\begin{equation}
\label{Eq:PathPro}
P[ \traj ] \propto p_0 ( \ket{\psi_0} ) \prod_{t=0}^{\tobs-\Delta t} u( |\psi_t\rangle \rightarrow |\psi_{t+\Delta t} \rangle),
\end{equation}
where $p_0 (|\psi_0\rangle)$ is the probability of observing the initial wavefunction and $u( |\psi_t\rangle \rightarrow |\psi_{t+\Delta t} \rangle)$ are the transition probabilities for each interval of time $\Delta t$. The transition probabilities represent the probability of waiting times between jumps multiplied by the probability for each jump, 
\begin{align}
u( |\psi_t\rangle \rightarrow |\psi_{t+\Delta t} \rangle ) =  1-\frac{\bra{\psi_t} \Gamma_n \op{L}^{\dagger}_n \op{L}_n \ket{\psi_t}}{r(\ket{\psi_t})}e^{- r(\ket{\psi_t}) \Delta t}
\end{align}
where $r(\ket{\psi_t})$ is the waiting time probability between jumps
\begin{equation}
r(\ket{\psi_t}) = \bra{\psi_t} \sum_n \Gamma_n \op{L}^{\dagger}_n \op{L}_n \ket{\psi_t},
\end{equation}
and the ratio in front of the exponential is the probabilty to make a jump due to the action of the $n$th Lindblad operator, both of which follow directly from Eq. ~\ref{Eq:SDE}. These transition probabilities have been shown to obey a differential Chapman-Kolmogorov equation and yield a Markovian stochastic process in the projective Hilbert space.~\cite{breuer1995stochastic}

We define the normalization of the path ensemble as the path partition function $Z(\tobs)$, which is obtained by integrating over all paths 
\begin{equation}
Z(\tobs) = \int \mathcal{D}[ \traj ] P[ \traj ],
\end{equation}
from which is clear that stochastic unraveling samples a real-time path integral, with probability measure $\mathcal{D}[ \traj ]$ for realizations over the Poisson random noise. The absence of a dynamical sign problem is due to the Markovian and weak system-bath coupling approximations. 
Observable quantities can be computed directly by averaging the time-dependent expectation value over the ensemble of trajectories
\begin{equation}
\peavg{\mathcal{O}(t)}  = \int \mathcal{D}[ \traj ] P[ \traj ] \qexp{\mathcal{O}}{t} 
\end{equation}
where the usual quantum operator expectation value at time $t \leq t_{obs}$ is averaged over the stochastic paths, denoted with $\langle \dots \rangle$. As a result of the detailed balance condition in Eq. \ref{eq:detbal}, the trajectories obeys microscopic reversibility as codified by the Crooks Fluctuation Theorem\cite{crooks1999entropy}. This result implies both the Jarzsynski equality\cite{jarzynski1997nonequilibrium} and the correct physical interpretation to the flow of energy into and out of the system through heat and work\cite{horowitz2012quantum,horowitz2013entropy}. 

While Eq. ~\ref{Eq:PathPro} denotes the total path probability, it is possible to only consider those trajectories that undergo a rare, or reactive event. To do this we define the probability of observing a rare event, $P_{AB} [\traj]$, in which the system begins in some quantum state  $A$ at time 0 and ends in some other quantum state $B$, at $\tobs$,
\begin{align}
&P_{AB} [ \traj ] \propto P[\traj] \bra{\psi_{0}} \op{h}_A \ket{\psi_{0}} \bra{\psi_{\tobs}} \op{h}_B \ket{\psi_{\tobs}},
\end{align}
where $\op{h}_{A(B)}$ is a projection operator for state $A$ ($B$).
The normalization of the path probability, $Z_{AB} (\tobs)$, and observables in this conditioned ensemble are computed as,
\begin{align}
&Z_{AB} (\tobs)=  \\ \nonumber
&\int D[\traj] P[\traj]\bra{\psi_{0}} \op{h}_A \ket{\psi_{0}} \bra{\psi_{\tobs}} \op{h}_B \ket{\psi_{\tobs}}
\end{align}
and,
\begin{equation}
\cpeavg{\mathcal{O}(t)}{AB} = \int \mathcal{D}[ \traj ] P_{AB}[ \traj ] \qexp{\mathcal{O}}{t},
\end{equation}
analogously as in the unconditioned path emsemble, and we adopt the subscript $AB$ on the brackets to denote an average in the reactive path ensemble. 
Here the semigroup property is requisite due to the dependence of the normalization on the physicality of individual trajectories. Though we specifically consider path ensembles conditioned on reactive events, this formalism is general, and can be used for conditioning on time extensive quantities as done with the $s$-ensemble and related techniques~\cite{budini2014fluctuating}.

If the probability of observing the transition form A to B in the unconstrained ensemble,  $P[ \traj ]$, is small, then accurately determining expectation values in the reactive path ensemble through brute force sampling will be difficult. One means to overcome such sampling problems is to use Transition path sampling (TPS) algorithms to sample  $P_{AB}[ \traj ]$ directly\cite{dellago2002transition,bolhuis2002transition}. 
Typically the most efficient Monte Carlo move for reactive path spaces is the so-called ``shooting move''\cite{dellago2002transition}. Shooting moves generate new trial trajectories by re-integrating the equation of motion forward and backward from some uniformly chosen intermediate time along the trajectory. If the integration of the trial trajectory uses the same equation of motion as that which defines the desired path ensemble, and the Monte Carlo procedure uses a symmetric change in the configuration about the intermediate time, the acceptance ratio is
\begin{align}
P_{\text{acc}}[\Psi^o \rightarrow \Psi^n] = \min \left\{1, \bra{\psi^n_{0}} \op{h}_A \ket{\psi^n_{0}}\bra{\psi^n_{\tobs}} \op{h}_B \ket{\psi^n_{\tobs}}\right \} 
\end{align}
where $\Psi^o$ and $\Psi^n$ are the old and new trajectories with their arguments suppressed for compactness, $P_{\text{acc}}$ is the acceptance probability for the Monte Carlo move, and the projection operators are evaluated at the end points of the new trajectory. Since the equation of motion for the quantum jump trajectory is stochastic, one-sided shooting can be done in order to increase the acceptance probability~\cite{dellago2002transition}. Here only the bias from the conditioning functional of $P_{AB}[\traj]$ appears due to the symmetry in the Monte Carlo moves. 

\subsection{Rate Constants}

Much like in the classical path ensemble formalism, a rate constant can be computed by a time derivative of the side-side correlation function, $C_{AB}(t)$,\cite{chandler1978statistical}
\begin{equation}
k (t) = \frac{d}{dt} C_{AB}(t)
\end{equation}
\noindent where
\begin{equation}
C_{AB} (t) = \frac{ \peavg{h_A[\psi_0] h_B [\psi_t] } }{\peavg{h_A[\psi_0]}} \, ,
\label{eq:qtcf}
\end{equation}
which is the conditional probability of the system being in state $B$ at time $t$, given the system started in state $A$ at time $t=0$. 
With the identification of the ensemble averages in Eq. \ref{eq:qtcf} as conditioned path partition functions it follows directly  just as it does with classical path ensembles\cite{dellago2002transition} that the rate constant is a time-derivative of a ratio of these conditioned path partition functions
\begin{equation}
k(t) = \frac{d}{dt} \frac{Z_{AB} (t)}{Z_A (t)},
\end{equation}
\noindent where
\begin{align}
Z_A (\tobs) &= \int D[\traj] P[\traj] \bra{\psi_{0}} \op{h}_A \ket{\psi_{0}}
\end{align}
is the reactant path partition function. The ratio of partition functions is computable by thermodynamic integration. By rewriting the ratio as an integral,
\begin{equation}
\ln \frac{Z_{AB}}{Z_A} = \int_0^B d\lambda \left( \frac{\partial \ln Z_{A\lambda}}{\partial \lambda} \right)
\end{equation}
the rate is identical to the reversible work to ``stretch'' the ensemble of trajectories from the reactant to product regions. While thermodynamic integration is one means to compute this reversible work, any other free energy method could be use analogously. To this end, umbrella sampling can be used to constrain trajectories beginning in region $A$ to end in overlapping intermediate regions $\lambda$ in the interval ranging from $A$ to $B$ and constructing a ``path free energy'' in this coordinate. Because the rate has been constructed as a ratio of path partition functions or likewise a difference of path free energies, the calculation is independent of path taken along the thermodynamic integration, hence \textit{a priori} knowledge of the reaction coordinate is unnecessary. 
The rate constant can then be computed either directly from the ratio of path partition functions or by computing the time-derivative. In the former case, one uses the identity at some steady-state time, \begin{equation}
k = \frac{1}{\tobs} \frac{Z_{AB} (\tobs)}{Z_A (\tobs)}.
\end{equation}
valid for $\tobs$ intermediate to the molecular timescale of a transition, $\tau_{\text{mol}}$, and the reaction timescale, $1/k$, ($\tau_{\text{mol}} < \tobs \ll 1/k$)
The rate constant is thus directly proportional to the ratio of path partition functions in this steady-state by the inverse of the steady-state time. Alternatively, the time derivative of this path partition function ratio can be computed by which the ratio is computed at a number of times and the slope, in the steady-state regime, is precisely the rate constant.

%%%%%%%%%%%%%%%%%%%%%%%%%%%%%%%%%%%%%%%%%%%%%%%%%%%%%%%%%%%%%%%%%%%%%%%%%%%%%
% DBA Section
%%%%%%%%%%%%%%%%%%%%%%%%%%%%%%%%%%%%%%%%%%%%%%%%%%%%%%%%%%%%%%%%%%%%%%%%%%%%%
\section{\label{sec:cond}Conditioned Ensembles}

\begin{figure}[t]
\centering
\includegraphics[width=8.5cm]{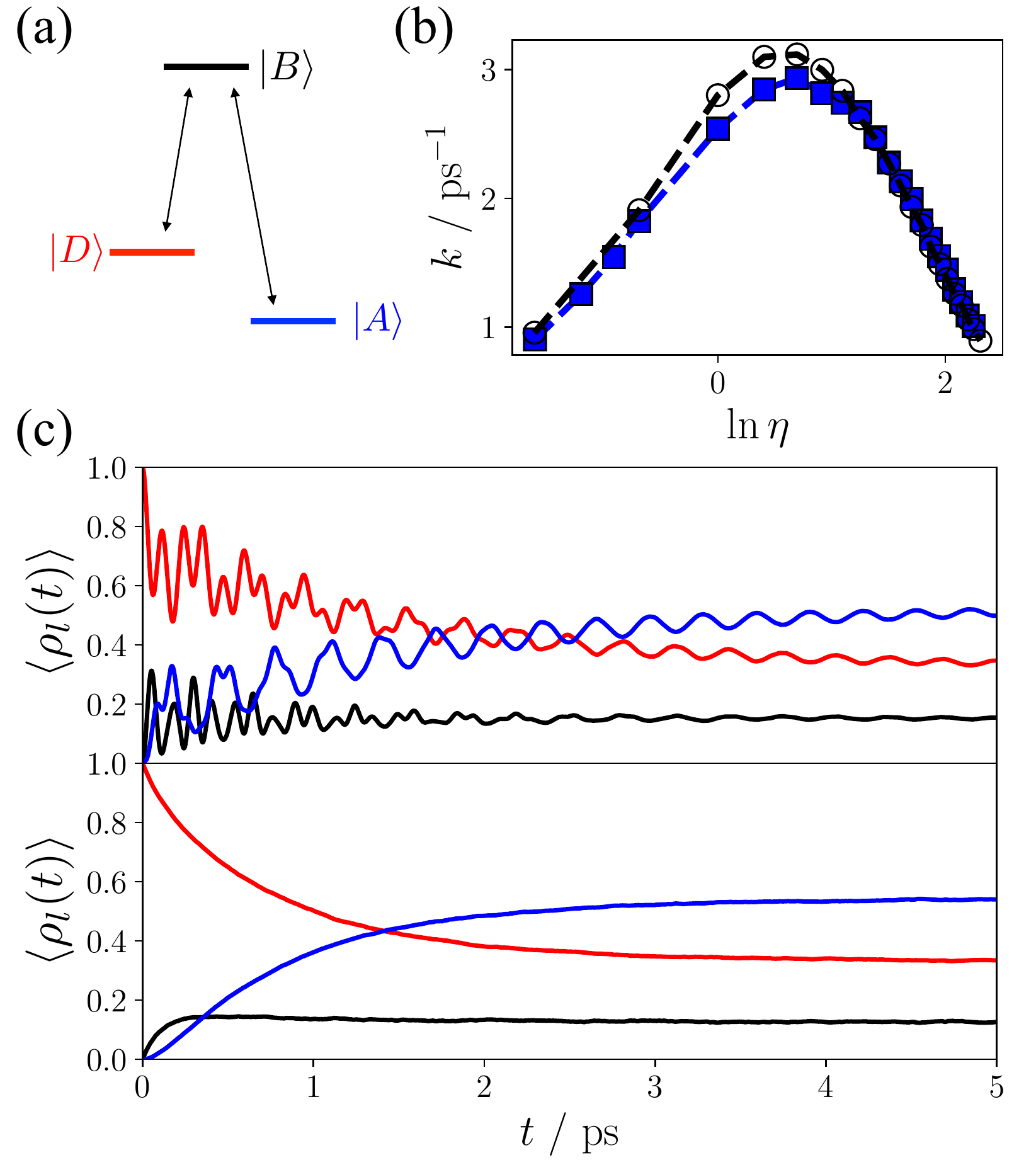}
\caption{Energy transfer dynamics in the DBA model. a) Schematic energy levels used in the study. b) Donor to acceptor energy transfer rate constants as a function of system-bath coupling, $\eta$, for the full PQME (open circles) and for the Lindblad PQME (blue squares). c) Population dynamics are shown for the donor (red), bridge (black), and acceptor (blue) sites for $\eta = 0.2$ (top panel) and $\eta = 9.0$ (bottom panel).}
\label{fig:dba_pops}
\end{figure}

In this section, we illustrate the utility of conditioned path ensembles for gaining mechanistic insight in open quantum dynamics. In particular, we show how conditioned ensembles build in correlations that elucidate the mechanistic details of specific rare events. Our work focuses on energy transfer dynamics in a donor-bridge-acceptor (DBA) system, schematically shown in Fig. \ref{fig:dba_pops}(a). This system was recently considered by Jang and co-workers who applied a novel quantum master equation, termed the polaron-transformed quantum master equation (PQME)\cite{jang2008theory,jang2011theory}, to a model three-level chromophoric system coupled to a bath\cite{jang2013coherent}. 
Depending on the strength of the coupling to the bath, the energy transport between the donor and acceptor states could follow from either a superexchange mechanism, in which an excitation initially localized on the donor state is transferred coherently to the acceptor state, or from a sequential hopping mechanism, in which the excitation is transferred incoherently through a barrier-crossing-like event to reach the acceptor state after passing through the intermediate bridge state. By increasing the coupling strength between the system and bath, one can observe a smooth transition between these mechanisms, which gives rise to an overall turnover in the rate of charge transfer.

The PQME method is able to treat a broad range of the system-bath coupling strength by making use of a small polaron transform to the original system bath model. This transformation incorporates the bath modes into the system Hamiltonian through a reorganization energy, which changes the site energies, and hopping integrals, which dampen the electronic coupling terms in the system Hamiltonian exponentially as the system-bath coupling strength increases. After the application of the small polaron transform, the system Hamiltonian becomes,
\begin{align}
\op{H}_{S} &= \sum_l \epsilon_l |l \rangle \langle l| + \sum_{l\neq l'} J_{ll'}  | l \rangle \langle l' |.
\end{align}
where  $l=D,B,A$, labels the donor, bridge and acceptor sites, $\epsilon_l$ are the site energies reduced by the reorganization energy, and $J_{ll'}$ are the inter-site couplings that are dressed by the polaron transform. In this model there are nonzero inter-site couplings between $D-B$ and $B-A$, but no direct coupling between $D-A$. A consequence of the polaron transform is that the form of $J_{ll'}$ depends on the system-bath coupling
\begin{align}
J_{ll'} = j_{ll'}e^{-\eta \lambda_r^2} 
\end{align}
where $ j_{ll'}$ are the bare inter-site couplings and are multiplied by an exponentially small term in the system-bath coupling strength, $\eta$, with temperature dependent prefactor 
\begin{equation}
\lambda_r^2 = \frac{\pi}{\eta} \int d\omega \, \mathcal{J} (\omega) \coth (\beta \hbar \omega / 2)
\end{equation}
which is a thermally weighted integral over the spectral density. Following Jang and coworkers,\cite{jang2013coherent} the spectral density is, using the convention of Eq. ~\ref{Eq:SpecDens}, taken to be of ohmic form
\begin{equation}
\mathcal{J} (\omega) = \frac{2}{\pi} \frac{\eta}{3!} \frac{\omega}{\omega_{c}^2} e^{-\omega/\omega_{c}},
\end{equation}
where $\omega_c$ is bath cutoff frequency. In principle, an inhomogeneous term arising from initial correlations between the system and bath modifies the system hamiltonian in a time dependent manner. However for the conditions we consider its effect is negligible, so we do not consider it in the following.

The resultant PQME is a weak-coupling master equation for the quasiparticle small polaron, interacting with the deformed environment, where the residual off-diagonal coupling to the bath is treated perturbatively~\cite{jang2011theory}.
In order to put the PQME into a quantum jump form, we must make two additional approximations to the equations of motion. First, we ignore non-secular terms that couple populations from coherences in the energy eigenbasis. Second, while the PQME is a time-local equation, it is non-Markovian in that the rates of transitions induced by the bath are time-dependent. In principle the Lindblad operators in the quantum jump equation can take time dependent forms, and as long as the rates are strictly positive the complete positivity of the density matrix will be preserved. However, we make a Markovian approximation and neglect this time-dependence. 

Given these approximations, we can construct Lindblad operators from the elements of a time-independent Redfield-like tensor.  As the rates of these operators obey detailed balance, it is most convent to express them in the energy eigenbasis. For population transfer between each pair of energy eigenstates, the Lindblad operators are 
\begin{equation}
\op{L}_{ij} = | \phi_i \rangle \langle  \phi_j|, \quad \Gamma_{ij} = \mathcal{R}_{iijj} 
\end{equation}
and the single dephasing operator is 
\begin{equation}
\op{L}_{\text{d}} = \sum_i \sqrt{\mathcal{R}_{iiii}} | \phi_i \rangle \langle \phi_i | 
\end{equation}
where we have absorbed  the dephasing rate into the dephasing operator, and so have $\Gamma_\mathrm{d} = 1$. The elements of the Redfield-like tensor follow directly from Jang et al and in the energy eigenbasis are,
\begin{equation}
\mathcal{R}_{iijj} =  \frac{1}{\hbar^2} \sum_{l\neq l'} \sum_{m \neq m'} J_{ll'} J_{mm'} \mathcal{F}_{ll',mm'}^{ii,jj} 
\end{equation}
whose kernel in our Markovian approximation is
\begin{align}
\label{Eq:Fthis}
\mathcal{F}_{ll',mm'}^{ii,jj} &=  \int_0^\infty dt \left(1-e^{-\mathcal{K}_{ll',mm'} (t)} \right)\times   \\
& \left (S_{ll',mm'}^{ij,jj} - \sum_{j'} S_{mm',ll'}^{ij,j'j'}  \right )+ \text{h.c.}, \nonumber
\end{align}
where $\mathcal{K}_{ll',mm'} (t) = (\delta_{lm}+ \delta_{l'm'} - \delta_{lm'} - \delta_{l'm}) C(t)$ and $\delta_{lm}$ is the Kronecker delta. The correlation function $C(t)$ is given by
\begin{equation}
C(t) = \int_0^{\infty} d\omega \mathcal{J}(\omega) [\coth(\beta \hbar \omega / 2) \cos (\omega t)-i\sin (\omega t)] \nonumber
\end{equation}
and the overlap factors, $S_{ll',mm'}^{ii,jj}$, coming from the change from the site to energy eigenbasis are given by
\begin{equation}
S_{ll',mm'}^{ij,j'j'} = \braket{\phi_i | m}\braket{m' | \phi_{j'}} \braket{\phi_{j'} | l} \braket{l' | \phi_j}  \nonumber
\end{equation}
and we employ $\text{h.c.}$ to refer to the Hermitian conjugate of the product of the previous terms in Eq.~\ref{Eq:Fthis}. 

Throughout we will use $\epsilon_B - \epsilon_D = 200$ cm$^{-1}$, $\epsilon_B - \epsilon_A = 200$ cm$^{-1}$, $j_{BD} = j_{BA} = 100$ cm$^{-1}$, and $\omega_c = 200$ cm$^{-1}$. With this equation of motion, and these parameters, we consider the dynamics of the system initially prepared in the donor state, $| D \rangle$. The donor state is energetically unfavored, and so relaxation mediated by the bath will lead to population transfer to the acceptor states. For the inter-site coupling strengths considered, the energy eigenstates are primarily localized on specific sites,  becoming exactly commensurate in the limit of large system bath coupling strength, $\eta$. For simplicity, we will label the energy eigenstates by $| \phi_l \rangle$, for the state primarily supported on site $l$,  and the corresponding state in the site basis with $| l \rangle$. 

\begin{figure}
\centering
\includegraphics[width=0.5\textwidth]{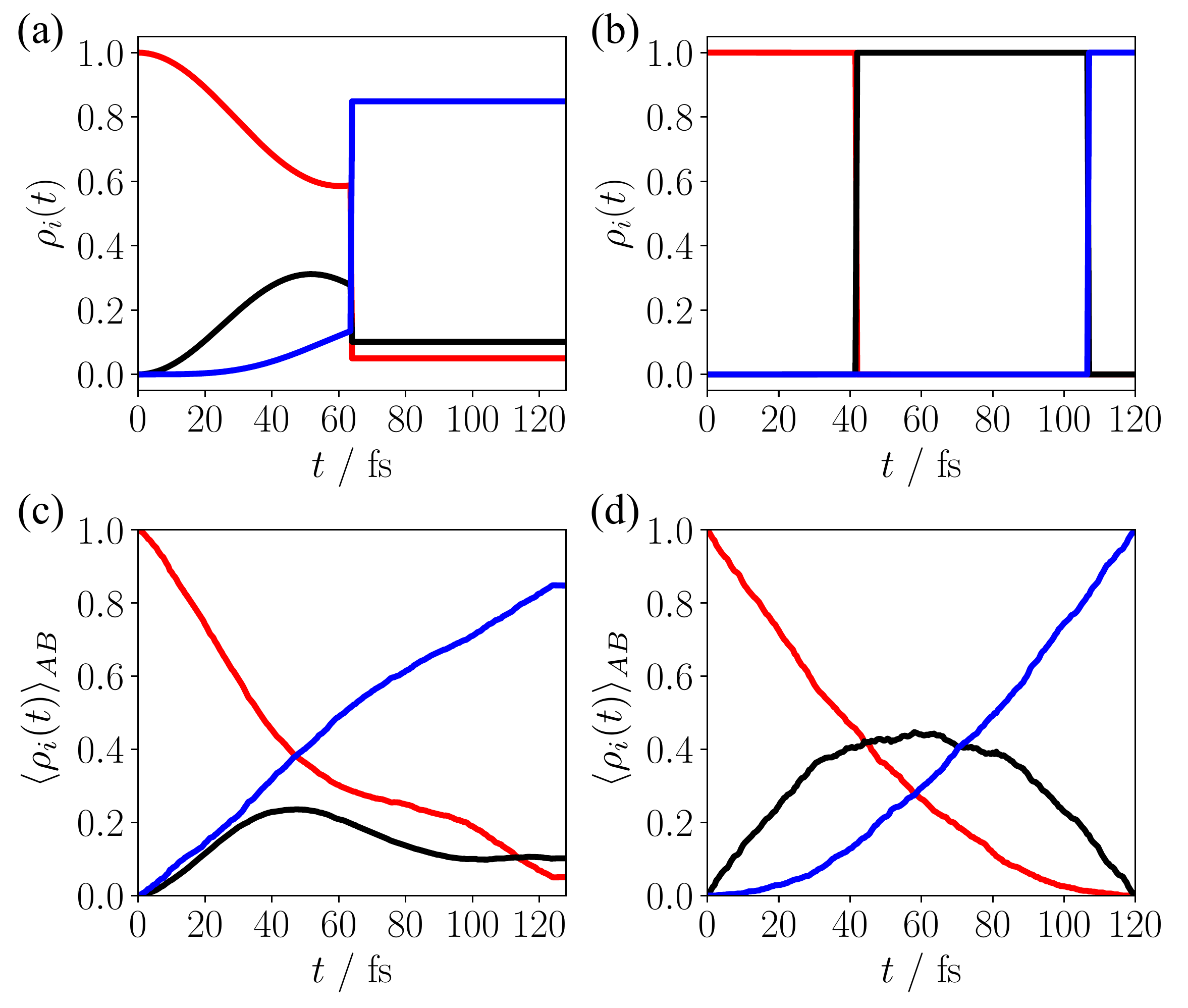}
\caption{Population dynamics in the reactive ensemble. Representative stochastic trajectories for the donor (red), bridge (black), and acceptor (blue) populations are shown for (a) for $\eta = 0.2$ and (b) for $\eta = 9.0$ and averaged populations are shown in (c) for $\eta = 0.2$ and (d) for $\eta = 9.0$. }
\label{fig:dba_cond_pops}
\end{figure}

Though the Lindbladization procedure described above invokes both the Markovian and secular approximation, the dynamics show quantitative agreement with the original simulations of Jang et. al.\cite{jang2013coherent} The rate constants computed from population dynamics, shown in Fig. \ref{fig:dba_pops}(b), are accurate across the whole range of system-bath coupling strengths exhibiting a maximum rate at $\eta = 2$, which agrees with the full PQME result. Example population dynamics computed from, an unconditioned ensemble, $\peavg{\rho_l(t)}$, where $\op{\rho}_l$ is the population operator, $\op{\rho}_l= \ket{l}\bra{l}$ for site $l=(D,B,A)$, exhibit the same qualitative changes from coherent dynamics at weak system-bath coupling to hopping dynamics at strong system-bath coupling. These results were accomplished with 40,000 trajectories for each $\eta$. As was noted in early applications of the PQME method,~\cite{jang2011theory} non-Markovian effects from the perspective of the non-transformed system Hamiltonian are treated in the system Hamiltonian to some degree by the PQME method due to incorporation of the bath modes from the small polaron transform. The deviation near the maximum stems from the secular approximation, which decouples additional transfer from coherences to the populations and slightly reduces the overall rate.

To study the mechanism of charge transport through trajectory analysis, we consider ensembles of trajectories conditioned on observing the system in the donor state at $t=0$ and in acceptor eigenstates at $t=\tobs$. These conditioned probabilities are computed in a reactive path ensemble with initial and final states given by the projectors
\begin{equation}
\op{h}_A = \ket{D}\bra{D} \quad \mathrm{and} \quad
\op{h}_B = \ket{\phi_A}\bra{\phi_A} 
\end{equation}
so that the system begins in the donor state, which is a superposition of energy eigenstates, undergoes dephasing and dissipation through the action of the bath, and ends in an energy eigenstate mostly localized in the acceptor state. Additionally, we take $\tobs=120$ fs, which is much shorter than the time for population decay from the donor state on average, as shown in Fig.~\ref{fig:dba_pops}(c), but long enough that the system builds up population in the acceptor eigenstate with high probability. 

\begin{figure}[b]
\centering
\includegraphics[width=\linewidth]{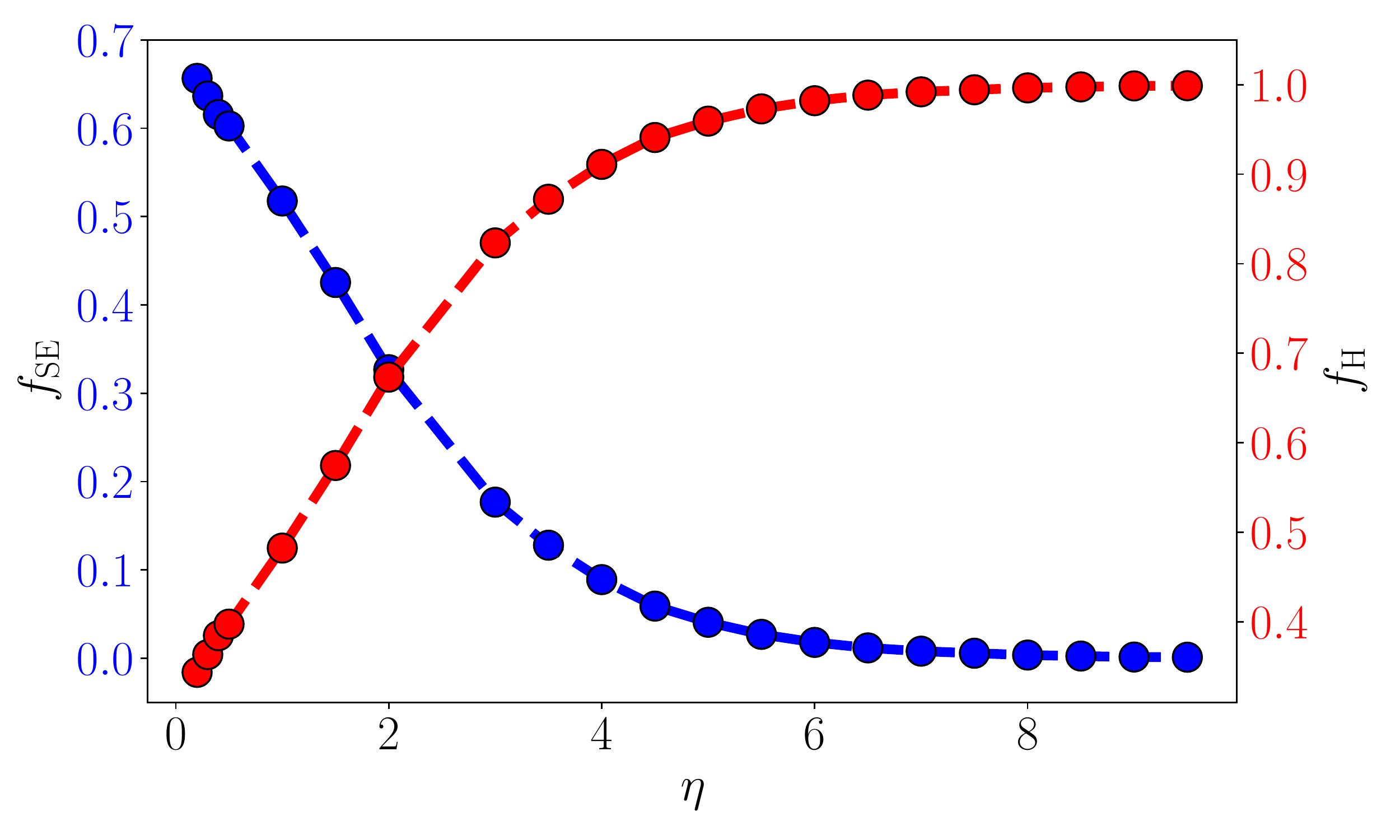}
\caption{The average number of mechanistic jumps per trajectory are shown as a function of $\eta$. The average number of superexchange jumps (dashed blue curve with blue circles) have values on the left y-axis and the average number of hopping jump sequences (dashed red curve with red circles) have values on the right y-axis. }
\label{fig:dba_d2a_jumps}
\end{figure}

Figures \ref{fig:dba_cond_pops}(a) and (b) show example quantum jump trajectories for $\eta = 0.2$ and $\eta = 9.0$, respectively. At weak coupling, the individual trajectories begin by undergoing Hamiltonian evolution with populations that are nearly identical to those in the unconditioned ensemble. After this initial delocalization through coherent dynamics, the system undergoes a quantum jump, which transfers population instantaneously between the eigenstates and gives rise to the decoherence apparent at long times in the averaged populations. Trajectories in the strong coupling regime are starkly different exhibiting no coherent evolution, due to the smaller inter-site coupling, and with quantum jumps transferring populations between the eigenstates. In the average populations, these quantum jumps result in exponential population transfer averaged populations, due to the exponential waiting time for the jump to occur. The short $\tobs$ consists largely of trajectories that have made donor-to-acceptor eigenstate transitions, but no reverse acceptor-to-donor transitions. 

The conditioned populations for $\eta = 0.2$ and $\eta = 9.0$ are shown in Fig. \ref{fig:dba_cond_pops}(c) and (d), respectively. In the weak coupling regime, where the superexchange mechanism dominates the transitions, the conditioned populations show a near direct transfer between the donor and acceptor states, while the dynamics in the bridge population remain nearly invariant to the conditioning relative to the unconditioned dynamics. At early times, the populations in the donor and bridge states rise at the same rate, and opposite to that of the acceptor states, which suggests that in this conditioned ensemble of trajectories the transfer follows the superexchange mechanism. In the strong coupling regime, where the hopping mechanism dominates, the conditioned populations show a sharp rise in the bridge state population followed by an increase in the acceptor state population. At short times the slopes now of the donor and bridge states are opposite one another, and at later times the slopes of the bridge and acceptor states are opposite. These features suggest that these trajectories primarily undergo hopping dynamics.

To verify this interpretation of the dynamics, we can directly resolve the bath operation that results in transfer from the donor to acceptor states in the individual quantum  jump trajectories. Specifically, superexhange trajectories are those in which the Lindblad operator that acts to localize the population on the acceptor eigenstate is either
$\op{L}_{DA}$ or $\op{L}_{BA}$ with no other population transfer jump occurring prior to these jumps. Using these operations ensures that transitions are made directly to the acceptor eigenstate either from the donor eigenstate or from the bridge eigenstate after coherent transfer of population to the bridge state. Hopping trajectories are similarly characterized with a Lindblad operator that localizes the population in the acceptor eigenstate directly from the bridge eigenstate, but only after first making a donor to bridge jump, $\op{L}_{BA}$ and $\op{L}_{DB}$, which offers the usual barrier crossing interpretation resulting from bath fluctuations.

With these characterizations, we can now directly test how each mechanism contributes to the dynamics of the density matrix over a range of $\eta$. Figure~\ref{fig:dba_d2a_jumps} shows the fraction of superexchange trajectories, $f_{\text{SE}}$, and the corresponding fraction of hopping trajectories, $f_{\text{H}}=1-f_{\text{SE}}$, that occur in the reaction path ensemble. 
 At weak system-bath coupling the majority of transfer events occurs via the superexchange mechanism, while at strong system-bath coupling the hopping mechanism is dominant. The decay of the fraction of superexchange jumps is exponential in the system bath coupling, which can be predicted by superexchange theory, due to the exponential decay of the inter-site coupling with increasing system-bath coupling in the polaron-transformed Hamiltonian. However, for all values of $\eta$ considered, the average rate of energy transfer is a combination of superexchange and hopping.
While superexchange theory predicts a monotonically decreasing rate, the rate of transfer via hopping is nonmonotonic, which is implied by the continued decrease in the overall rate in  Fig. \ref{fig:dba_pops}(b), in the strong coupling regime where the mechanism is dominated by hopping transitions. This nonmonotonic behavior is the result of self-trapping, which decreases the rate at large values of $\eta$.

%%%%%%%%%%%%%%%%%%%%%%%%%%%%%%%%%%%%%%%%%%%%%%%%%%%%%%%%%%%%%%%%%%%%%%%%%%%%%
% PCET Section
%%%%%%%%%%%%%%%%%%%%%%%%%%%%%%%%%%%%%%%%%%%%%%%%%%%%%%%%%%%%%%%%%%%%%%%%%%%%%
\section{\label{sec:comm}Committor Analysis}

In the context of photo-induced nonadiabatic dynamics, the rate of a event is often less important than its associated yield. The yield of such a process depends on how the dynamics of a specific chemical system favors forming the product state over relaxing back to the reactant state.
In the context of chemical reactions this manifests itself in the chemical selectivity. In this section, we show how path ensembles can be used to understand this selectivity by studying the dynamics of a proton-coupled electron transfer (PCET) model developed by Hammes-Schiffer and co-workers\cite{venkataraman2009photoinduced}. In particular, we show how stochastic unraveling can be used to interrogate the relaxation mechanisms that determine quantum yield following photoexcitation using a generalization of commitment analysis in corporate the commitment to different potential product states. Understanding the mechanism of yields is of broad importance to understanding a number of chemical reactions in photochemistry such as photoisomerization reactions~\cite{hahn2002ultrafast,balzer2005modeling} and other relaxation phenomena like hot carrier generation~\cite{deotare2015nanoscale}.

\begin{figure}
\centering
\includegraphics[width=\linewidth]{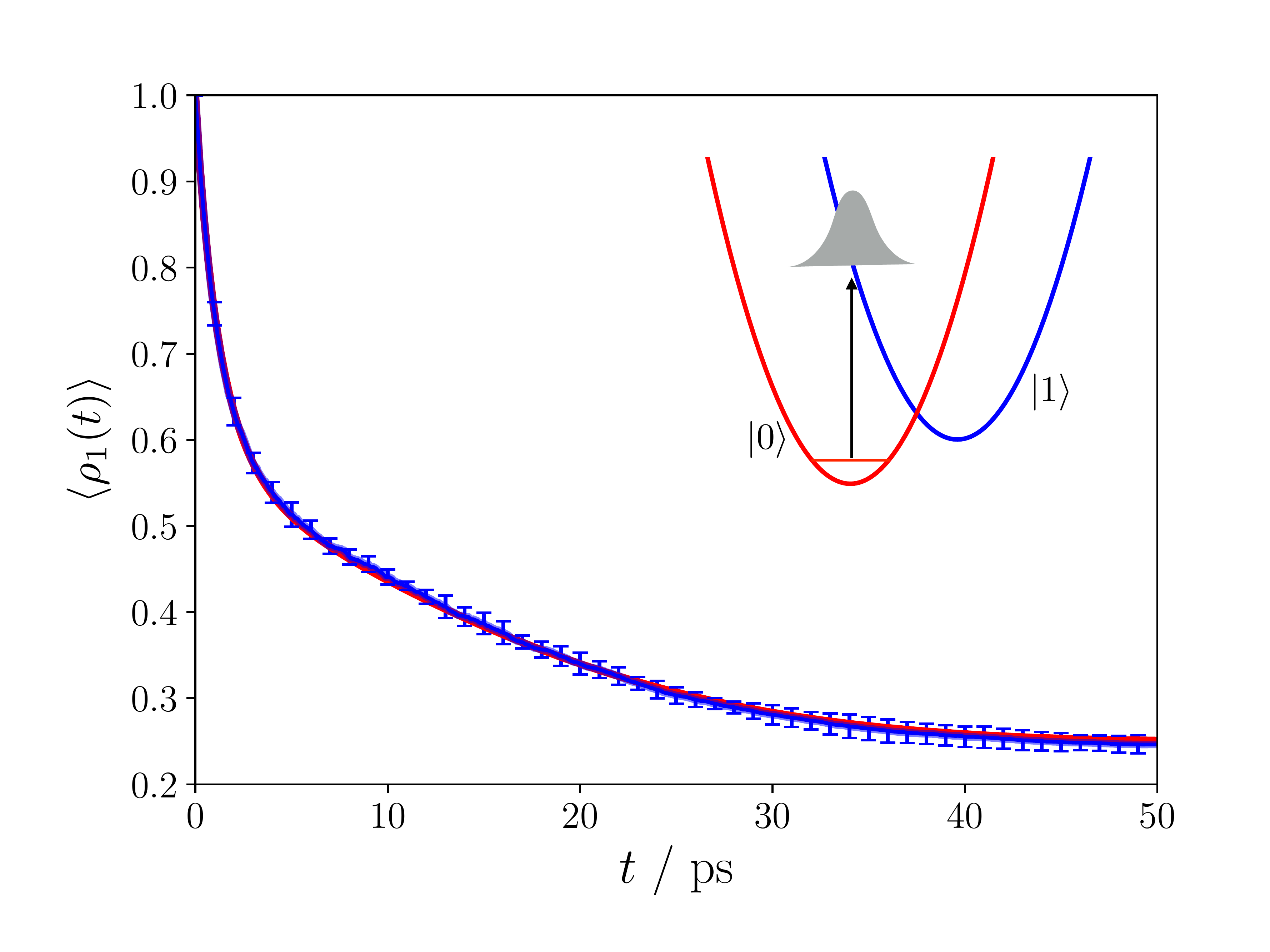}
\caption{Population dynamics of PCET model from RDM simulation (red) and from stochastic unraveling (blue). Error bars are computed from block averaging and represent a 95 percent confidence interval. The inset shows the potential energy surfaces of the acceptor (red parabola) and the donor (blue parabola) states.}
\label{fig:pcet_pops}
\end{figure}

The model we study (model A from Ref. \onlinecite{venkataraman2009photoinduced}) describes the photoinduced PCET for a system with electronic energy bias coupled to a bath. The system is composed of two harmonic oscillators, depicted in Fig. \ref{fig:pcet_pops}, coupled to a  harmonic oscillator bath. The system is strongly coupled to this bath, so the small polaron transform is again utilized to ensure the accuracy of a weak coupling perturbation theory. In this polaron-transformed frame, the system Hamiltonian is, 
\begin{align}
&\op{H}_S =  -\frac{\hbar^2}{2 m} \frac{\partial^2}{\partial \hat{q}^2} + \sum_{l=0,1}  U_l(\op{q}) \ket{l}\bra{l}
\end{align}
where $\hat{q}$ is the proton coordinate with mass $m$, $l$ labels the donor, $\ket{1}\bra{1}$, and acceptor, $\ket{0}\bra{0}$, electronic states, each with an
associated harmonic potential energy, $U_l(\op{q})$
$$
U_l(\op{q}) = \frac{1}{2} m \omega_l^2 (\op{q}-q_l)^2 + \epsilon_l
$$
where, $\epsilon_l$ is the potential energy minimum, $q_l$ its equilibrium position, and $\omega_l$ its characteristic frequency.
In this model, the system-bath coupling is treated in the electronic coupling, so that the bath serves to localize the excited electron and reduce the rate of electronic oscillation arising from off-diagonal coupling in the original system Hamiltonian. The electronic coupling can then be treated perturbatively with secular Redfield theory. Thus in this model electron transfer occurs because of bath fluctuations that temporarily permit the coherent electron transfer.

The resulting Lindblad operators are population transfer operators between the vibrational states on different electronic states, 
\begin{equation}
\op{L}_{l n,l' n'} = \ket{l}\ket{n}\bra{n'}\bra{l'}, \quad \Gamma_{l i,1 j} = \mathcal{R}_{l n, l' n'}
\end{equation}
where the pair $l n$, label the electronic state $l=\ket{0},\ket{1}$ and $n$, the vibrational eigenstate. There is one dephasing operator that is just the unit operator for the donor state
\begin{equation}
\op{L}_{\text{d}} = \sum_n \sqrt{G_{1n}} \ket{1}\ket{n}\bra{n}\bra{1}  .
\end{equation}
weighted by the rate $\sqrt{G_{1n}}$, so that $\Gamma_{\text{d}}=1$. The population transfer rates are given by a Fourier-transform of the bath-correlation function formally expressed for acceptor to donor transitions and donor to acceptor transitions, respectively, as
\begin{equation}
\mathcal{R}_{ln,l' n'} = \frac{1}{\hbar^2} |V_{ll'}|^2 |F_{n n'}|^2  \int_{-\infty}^{\infty} dt \, e^{i (E^{l}_n-E^{l'}_{n'})t/\hbar} M(t) 
\end{equation}
where $V_{ll'}$ is the electronic coupling matrix element which is nonzero only for $l \ne l'$, $F_{n n'}$ is the Franck-Condon overlap factor between the vibrational states on different electronic states, $F_{n n'} = \langle 0 |  \langle n | n' \rangle | 1 \rangle$, and $E^{l}_n$ is the energy of the $n$th vibrational state of the $l$th electronic state, and $M(t)$ is the thermally averaged, polaron transformed, bath correlation function. The elements of this tensor give the rate of transfer between the vibrational states of each electronic state. The dephasing rates are given by
\begin{equation}
G_{1n} = \sum_{n'} \mathcal{R}_{1n,0 n'} 
\end{equation}
as the bath is only coupled to the donor electronic state, only the coherences of the donor state undergo dephasing.
Following Ref. \onlinecite{venkataraman2009photoinduced} the bath correlation function is computed using a high-temperature approximation, 
\begin{equation}
M(t) \approx \exp \left( -\frac{\lambda_s t^2}{\hbar^2 \beta} - \frac{i t \lambda_s}{\hbar} \right) \, ,
\end{equation}
where $\lambda_s$ is the reorganization energy. Given the form of the Lindblad operators and their associated rates, population transfer only occurs between vibrational states of different electronic states, with an average dissipation roughly given by $\lambda_s$.  As the original dynamics were simulated with the secular approximation, the Lindblad master equation we employ gives equivalent dynamics, just in a different representation.

We consider dynamics following a vertical excitation of the ground vibrational eigenstate of the acceptor into the donor electronic state. The subsequent initial condition, $\ket{\psi^v_0}$, is illustrated in Fig.~\ref{fig:pcet_pops} and is given by
\begin{equation}
\ket{\psi^v_0} = \sum_{i}  c_n | i \rangle | 1 \rangle 
\end{equation} 
where the coefficient $c_n =  \langle 0| \langle 0 | n \rangle | 1 \rangle$ is the vibrational overlap factor of the 0th vibrational state of electronic state 0, with the $i$th vibrational state of electronic state 1. Throughout this section we use $\Delta \epsilon = \epsilon_1 - \epsilon_0 = 1$ eV, so that the acceptor state is energetically preferred, $\omega_0=\omega_1= 3000$ cm$^{-1}$, and $q_0= -0.5$ \AA, $q_1=0$ \AA, The electronic coupling is taken to be $V_{01}=0.03 $ eV, $m=1$ amu, the mass of a hydrogen atom, the temperature is $T=300$ K and the reorganization energy is $\lambda_s=0.892$ eV. 
For these parameters and initial condition, we find we can truncate the Hilbert space to include only the lowest 30 vibrational levels in each electronic state. The population dynamics in the donor state, $\peavg{\rho_1 (t)}$ where $\op{\rho}_1 = \sum_n \ket{1}\ket{n}\bra{n}\bra{1}$, following this vertical excitation are compared between the reduced density matrix formalism and simulation with stochastic unraveling in Fig. \ref{fig:pcet_pops}. With 40,000 trajectories the population dynamics are well-converged and exhibit the same dynamical features. With these choices of parameters, following a fast initial relaxation aided by the large Franck-Condon overlap for high energy states, a metastable population forms at intermediate times relative to the equilibrium distribution in which the donor-state population is negligible. This metastable state is due to a branching process that occurs during the vibrational relaxation that splits population into the donor and acceptor states, resulting in an enhancement of population in the donor state, 0.3, over its equilibrium value, essentially 0.0. Using trajectory analysis we can clarify the mechanism by which this branching occurs and thus understand what bath fluctuations give rise to a preferential population of the donor state over the acceptor state.

\begin{figure}
\centering
\includegraphics[width=8.5cm]{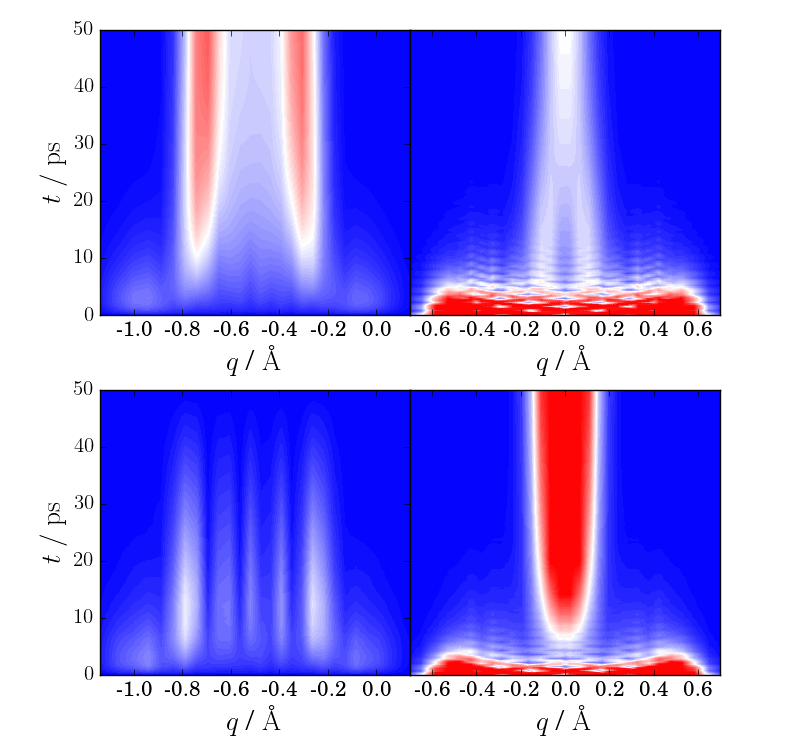}
\caption{Projections of the wavepackets onto the position basis ($q$) in the acceptor state (left column) and the donor state (right column) for the unconditioned path ensemble (top row) and the conditioned path ensemble (bottom row) as described in the text. Positions of high wavepacket probability are in red and near zero are blue. All plots use a single color range.}
\label{fig:pcet_q}
\end{figure}

To study the mechanism of preferential relaxation into the donor state, we define the reactive path ensemble for this model as
\begin{equation}
\op{h}_A = \ket{\psi^v_0}\bra{\psi^v_0} \quad \mathrm{and} \quad
\op{h}_B = \ket{1}\ket{0}\bra{0}\bra{1} 
\end{equation}
where the vertically excited initial condition is taken as the reactant and ground vibrational level of the donor as the product, and consider $\tobs=50$ ps which is long enough to observe initial relaxation to the ground vibrational state of the donor, but shorter than the characteristic time to thermally transfer population from the donor, over the potential barrier to the acceptor state. 
As was noted in Ref. \onlinecite{venkataraman2009photoinduced}, the projections of the wavepacket onto the coordinate basis shows the relaxation into each minima. 
We have computed analogous wavepacket projections which are constructed by $\chi(q,t) =  \braket{ q | \psi_t}$ where $\ket{q}$ is an eigenvector of the position operator $\op{q}$ and compared them with those averaged in the reactive path ensemble $\langle |\chi(q) |^2 \rangle_{AB}$ to those in unconditioned ensemble, $\langle |\chi(q) |^2 \rangle$. 
These are shown in Fig. \ref{fig:pcet_q}, where the normalization is computed for the both ensembles by ensuring wavefunction normalization at $t=0$. Figure \ref{fig:pcet_q} shows how the conditioned wavepacket begins branching from the unconditioned wavepacket, at roughly 10 ps seemingly commiting to either the donor and acceptor state after undergoing an initial dephasing which damped the oscillations in the donor state. 

While the averaged dynamics illustrate correlations between early time wavepacket motion and eventual localization in the donor or acceptor states, specific causal relationships and mechanistic information cannot be determined from them alone. In order to clarify the specific mechanism by which relaxation preferentially localizes in the donor state we have performed a committor analysis~\cite{dellago2002transition,bolhuis2002transition,peters2016reaction}. 
For each trajectory within the reactive path ensemble, we compute the probability, $p_B(t)$, that a given state of the system at some intermediate time $0<t<\tobs$ commits to the donor state. This is computed by averaging the fraction of trajectories that localizes in the donor state, integrated from the common intermediate state. 

\begin{figure}[b]
\centering
\includegraphics[width=\linewidth]{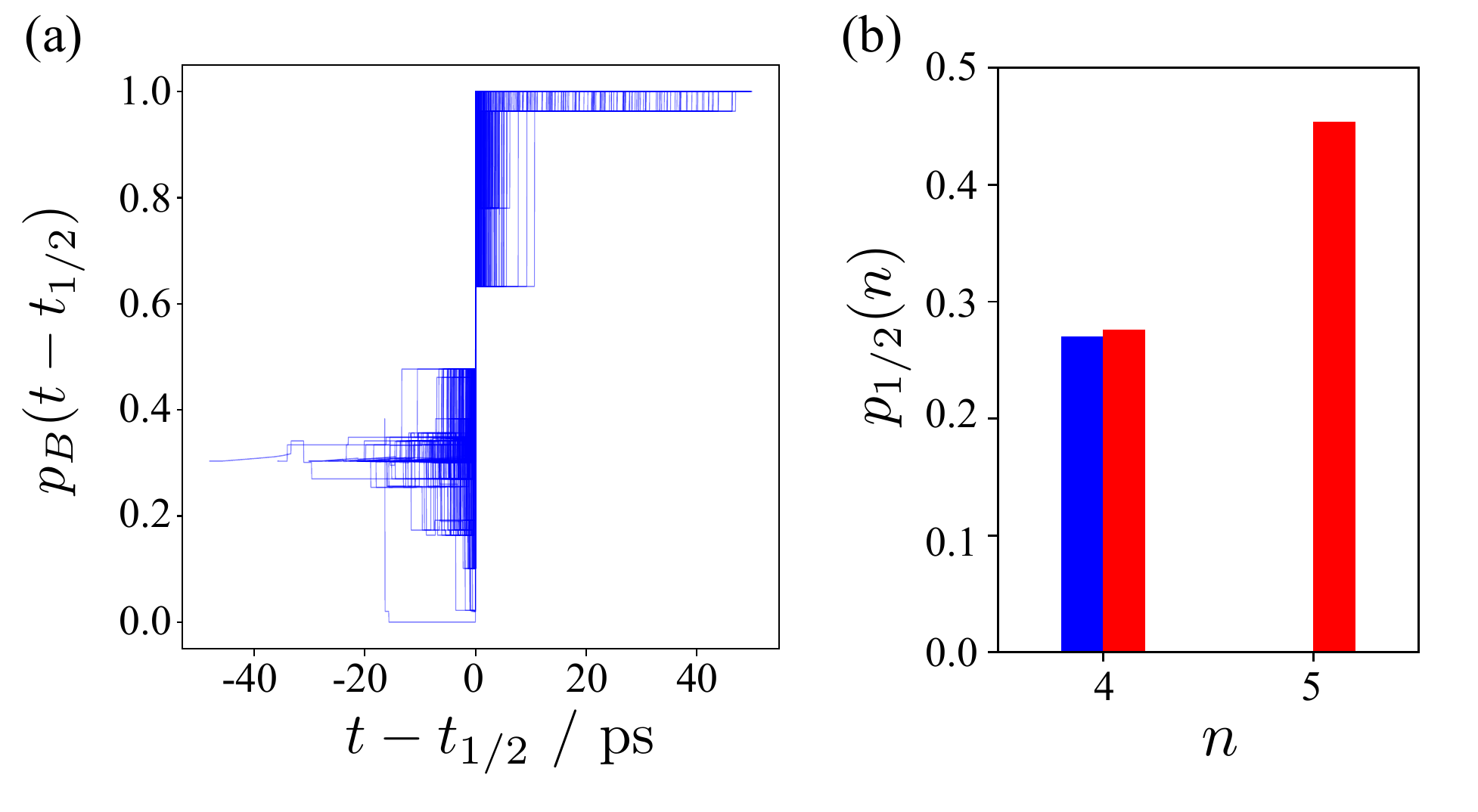}
\caption{Commitment probabilities as a function of time along each trajectory are shown in a). Each probability is shifted in time by $t_{1/2}$ the time when the commitment probability jumps to greater than 1/2. The fraction ($p_{1/2} (n)$) of configurations at $t_{1/2}$ in the $n^{\text{th}}$ vibrational state of the acceptor state (red bars) and the donor state (blue bars) is shown in b).}
\label{fig:pcet_commit_probs}
\end{figure}

Figure \ref{fig:pcet_commit_probs}(a) shows the commitment probabilities along all of the reactive trajectories taken from the unconditioned ensemble. At the initial time of each trajectory the commitment probability is the same and equal to the unconditioned yield of the reaction. which in this case is 0.3. Over the trajectory time, $p_B(t)$ changes as each trajectory begins to jump into different vibrational eigenstates that are more or less likely to localize in the donor state. At long times, $p_B(t)$ approaches 1, as required for a member of the reactive path ensemble. For each trajectory there is a unique time, $t_{1/2}$, where the commitment probability jumps above 1/2. The ensemble of configurations defined by the state of the system at $t=t_{1/2}$ are members of a transition state ensemble. By understanding the commonalities of trajectories in this ensemble, we can identify the required dynamical fluctuation for ending in the donor state.

\begin{figure}[t]
\centering
\includegraphics[width=\linewidth]{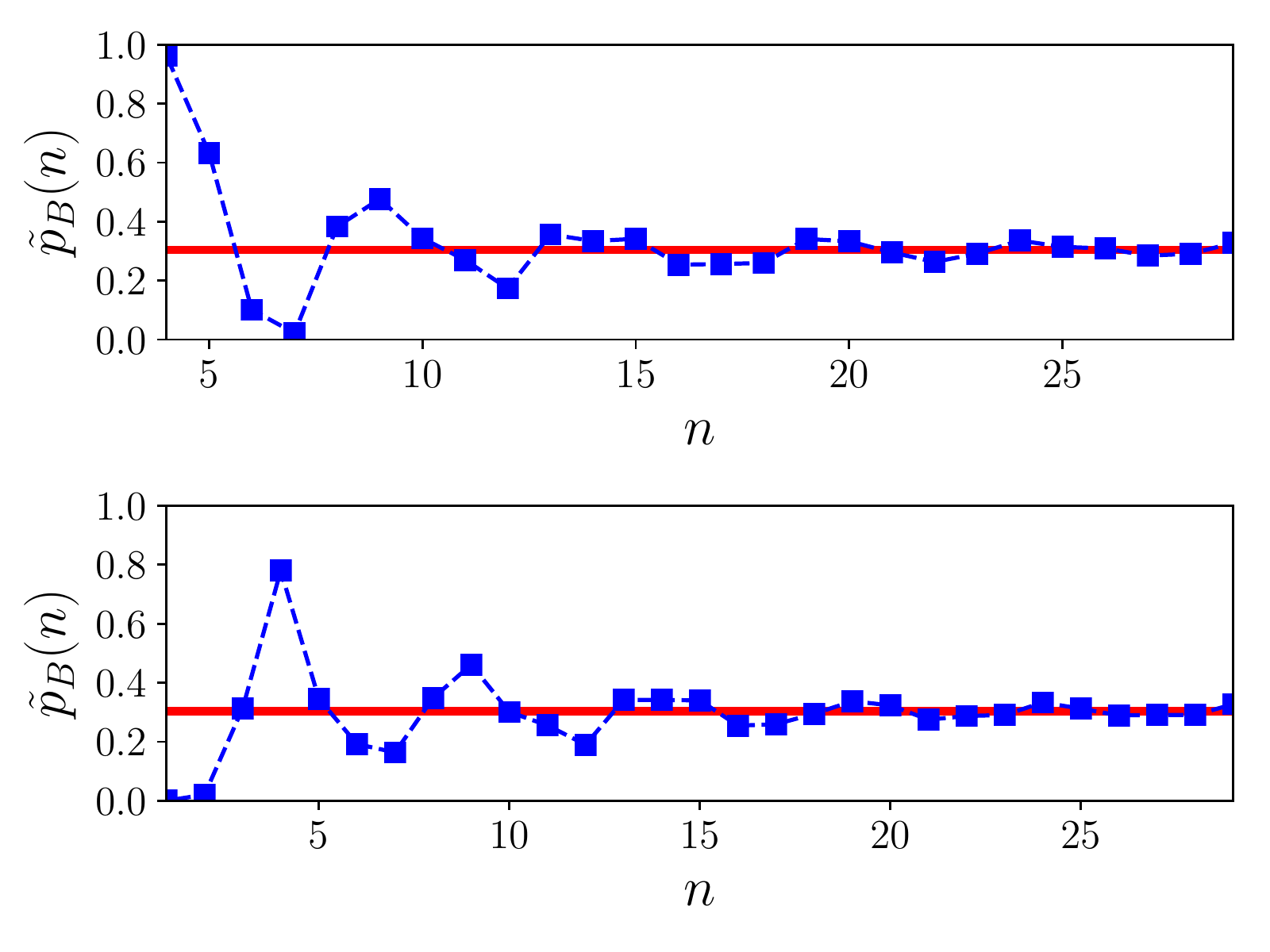}
\caption{Commitment probabilities for initialization in each vibrational state of the acceptor (top panel) and the donor (bottom panel). The red line is the commitment probability from the unconditioned path ensemble.}
\label{fig:pcet_commit}
\end{figure}

By analyzing the transition state ensemble, we have found that there are specific vibrational relaxation pathways that contribute the yield of the donor state. We have identified these pathways by computing the probability, $p_{1/2}(n)$, that members of the transition state ensemble reside in a particular vibrational state of the donor or acceptor,
\begin{align}
p_{1/2}(n) = \int \mathcal{D} [\traj]  P_{AB} [ \traj ] \nonumber \\
\times \delta (n - \langle \psi_{t_{1/2}}| \op{n} | \psi_{t_{1/2}} \rangle) 
\end{align}
where the average is over the reactive ensemble, $\op{n}=\sum_l |l\rangle | n\rangle \langle n | \langle l |  $ and the time is taken as the commitment time. Figure \ref{fig:pcet_commit_probs} (b) shows the fraction of vibrational states in the transition state ensemble, which has support over only 3 states, the 4th and 5th vibrational state of the acceptor and the 4th vibrational state of the donor. These states are greater in energy than the ground vibrational state of the donor state by either twice the solvent reorganization energy in the case of the donor state or just the solvent reorganization energy in the acceptor state.

To understand the importance of the reorganization energy in determining the commitment probability we computed the commitment probability, $\tilde{p}_B(n)$, for starting in a given vibrational state on either electronic states, unconditioned on being a member of the reactive path ensemble. This is shown in shown in Fig.~\ref{fig:pcet_commit}. As a function of the vibrational state, the commitment probability oscillates around the unconditioned value of 0.3. The oscillations in these commitment probabilities have a period of nearly 2 times the reorganization energy. Comparing this to the transition rates computed from the $\Gamma_{ij}$'s it is clear that the average dissipation incurred by a jump is given by the solvent reorganization energy, and the bottleneck to localizing in the donor state is passing through specific vibration levels whose energy the bath can most effectively dissipate. Hence, the statistics of the dissipation for each jump has a determining impact on the commitment probability and subsequently the quantum yield. Within this small polaron framework, this result suggests that engineering the reorganization energy by changing the solvent could be used to enhance the yield of photo-induced PCET.

%%%%%%%%%%%%%%%%%%%%%%%%%%%%%%%%%%%%%%%%%%%%%%%%%%%%%%%%%%%%%%%%%%%%%%%%%%%%%
% Rates Section
%%%%%%%%%%%%%%%%%%%%%%%%%%%%%%%%%%%%%%%%%%%%%%%%%%%%%%%%%%%%%%%%%%%%%%%%%%%%%
\section{\label{sec:rates}Evaluation of Rate Constants}

Computing rate constants can often be a challenging endeavor, especially for systems with rare events that control the rate process. In those systems, simple rate theories like Transition State Theory\cite{eyring1935activated,wigner1938transition} (TST) are relied upon due to their ease of implementation, however, such theories often break down for systems in condensed phases due to entropic effects and recrossing events that are excluded in the theory. Furthermore the application of many simple theories requires \textit{a priori} detailed knowledge of the mechanism, which can be elusive in complex condensed-phase systems. In this section, we utilize the path ensemble formalism to compute a rate constant in a model system with rare barrier crossing transitions. 

\begin{figure}[t]
\centering
\includegraphics[width=8.5cm]{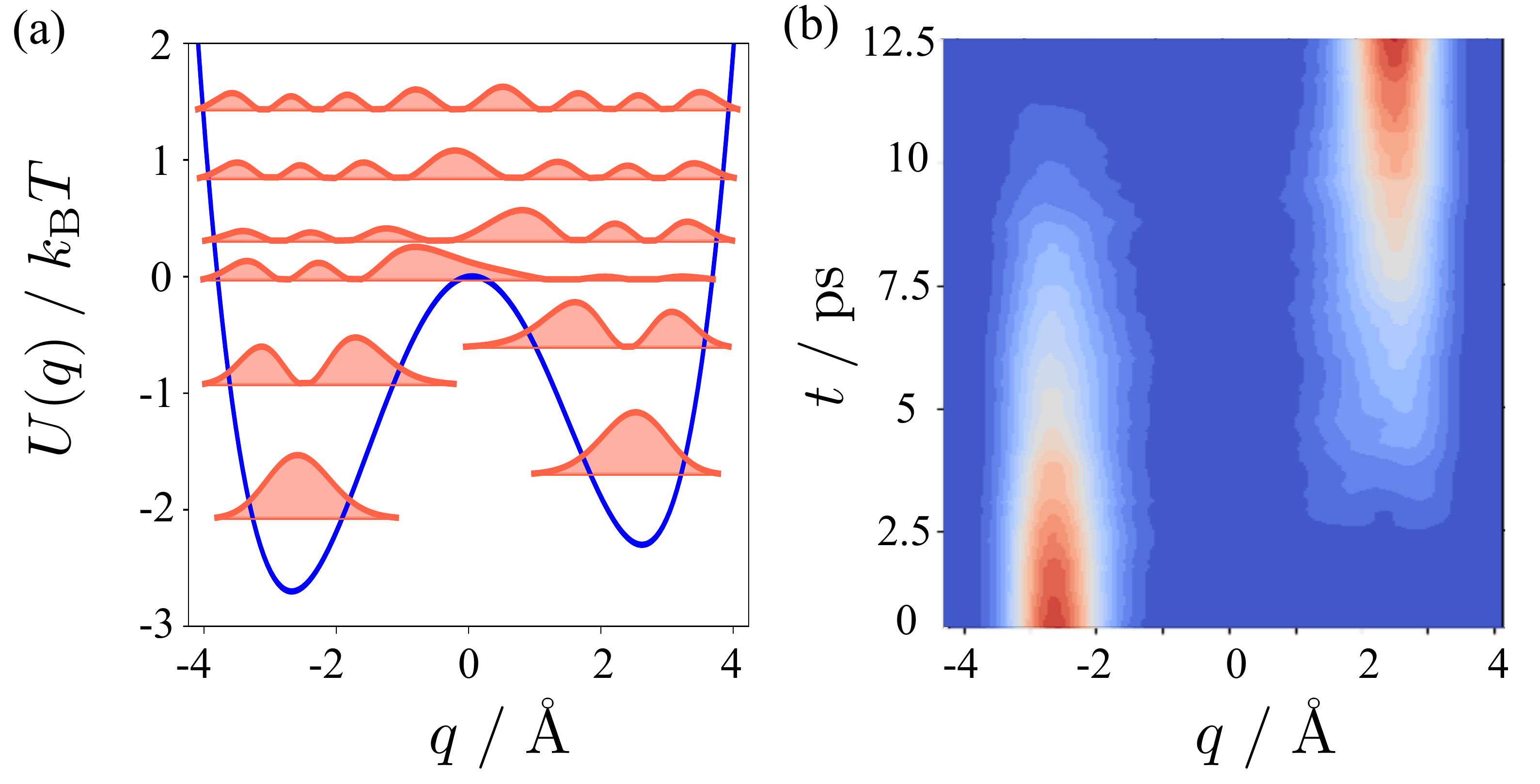}
\caption{Model for thermally activated barrier crossing. (a)The quartic potential used as a function of the position (blue) with its associated eigenstate wavefunctions (red filled curves). (b) The average wavepacket conditioned on beginning in the left well and evolving to the right.}
\label{fig:quartic_model}
\end{figure}

The model in question has a system Hamiltonian
\begin{equation}
\hat{H}_s = -\frac{\hbar^2}{2 m} \frac{\partial^2}{\partial \op{q}^2} + U(\op{q}).
\end{equation}
\noindent The potential (depicted in Fig. \ref{fig:quartic_model}), $U(\op{q})$, has a quartic polynomial form
\begin{equation}
U(\op{q}) = a \op{q}^4 - b \op{q}^2 + \epsilon \op{q},
\end{equation}
where $\op{q}$ is the position operator. The first two terms in the potential are necessary for producing a symmetric double-well potential, while the linear term induces a bias to one well that breaks the symmetry, a requirement for obtaining eigenstates that are localized to each well. In units of $\hbar = 1$ we have taken the mass of the particle to be $m=1$ and $\beta = 2 \times 10^3$ with dimensionless potential parameters $a=0.02$ $\kB T$, $b=-1.0$ $1\kB T$, and $\epsilon=0.2$ $\kB T$.
 The eigenstates are found using the sinc-function discrete variable representation (DVR) basis of Colbert and Miller\cite{colbert1992novel}. The DVR grid was uniformly spaced over a range $q\in [-8,8]$ with a distance $\Delta q = 0.05$ \AA. Despite the large basis set required for converging the eigenstates, only the lowest 10 eigenstates, which are labeled in energy-ascending order from 0 to 9, were needed in propagating the dynamics. 

We construct the Lindblad operators using a weak coupling secular Redfield theory where for each energy eigenstate pair $\ket{\phi_i}$ and $\ket{\phi_j}$ we have population transfer operators given by
\begin{equation}
\op{L}_{ij} = \ket{\phi_i}\bra{\phi_j}
\end{equation}
and rates, $\Gamma_{ij}$, given by
\begin{align}
\Gamma_{ij} = \frac{1}{\pi} \int_0^{\infty} dt e^{-i \omega_{ij} t} \int_0^{\infty} d\omega \mathcal{J}(\omega) &[\coth (\beta \hbar \omega /2) \cos (\omega t)  \nonumber \\
&- i \sin (\omega t)]
\end{align}
where 
the spectral density, $\mathcal{J}(\omega)$, has an Ohmic form with an exponential cutoff 
$$
\mathcal{J}(\omega) = \eta \omega e^{-\omega / \omega_c}
$$
with a coupling strength of $\eta = 0.01$ and cutoff frequency $\omega_c = (E_2 - E_0) / \hbar$. With these parameters the system is very weakly coupled to the bath, so secular Redfield theory is accurate, and the cutoff frequency is chosen to induce vibrational relaxation in each well of the quartic potential. Transitions between the wells will primarily occur as a result of barrier crossing, as is shown in the average wave-packet dynamics in the reactive ensemble in Fig. \ref{fig:quartic_model}(b). This trajectory illustrates directly the importance of tunneling in the model, as an initially localized wavepacket in the reactant state transfers to the product state without having much support present in the barrier region. Since within the secular approximation, populations and coherences are decoupled, for simplicity we neglect dephasing operations without loss of generality. 

We define a reactive path ensemble for transitions between the left and right well, as defined by
\begin{align}
\hat{h}_A &= | \phi_0 \rangle \langle \phi_0 | + |\phi_2 \rangle \langle \phi_2 | \quad \hat{h}_B = |\phi_1 \rangle \langle \phi_1 |,
\end{align}
which represent projectors for the lowest two eigenstates of the left well and the lowest eigenstate of the right well. The initial condition was a thermal distribution restricted to the reactant region,
\begin{equation}
\psi_0 = \sqrt{\frac{e^{- \beta E_0}}{Z}} |\phi_0 \rangle  + \sqrt{\frac{e^{- \beta E_2}}{Z}} |\phi_2 \rangle 
\end{equation}
where $Z=e^{- \beta E_0}+e^{- \beta E_2}$. The rate constant from population dynamics, $k^\mathrm{pop}$, is given by the time-derivative of the population in the product state, 
\begin{equation}
k^\mathrm{pop} = \frac{d \langle h_B(t) \rangle}{dt}\, ,
\end{equation}
and when evaluated in the steady-state regime, the rate of the transition is estimated to be $k^\mathrm{pop}  = 0.0106$ ns$^{-1}$. 

The rate constant was also computed via TPS, as outlined in Sec. \ref{sec:theory} C. Specifically, the ratio of path partition functions was estimated using umbrella sampling~\cite{torrie1977nonphysical}. We employed umbrella potentials of the form of hard walls to constrain the $B$-region of the trajectories using overlapping indicator functions of different eigenstates, denoted by $\op{h}_{\lambda}$, that were observed along typical transition paths. These umbrella potentials constrained the final wavefunction to be projected into an eigenstate contained in $\lambda$ and by using overlapping indicator functions. The full path partition function could be reconstructed as a function of $\lambda$ using histogram reweighted techniques~\cite{kumar1992weighted,shirts2008statistically}. 

\begin{figure}
\centering
\includegraphics[width=8.5cm]{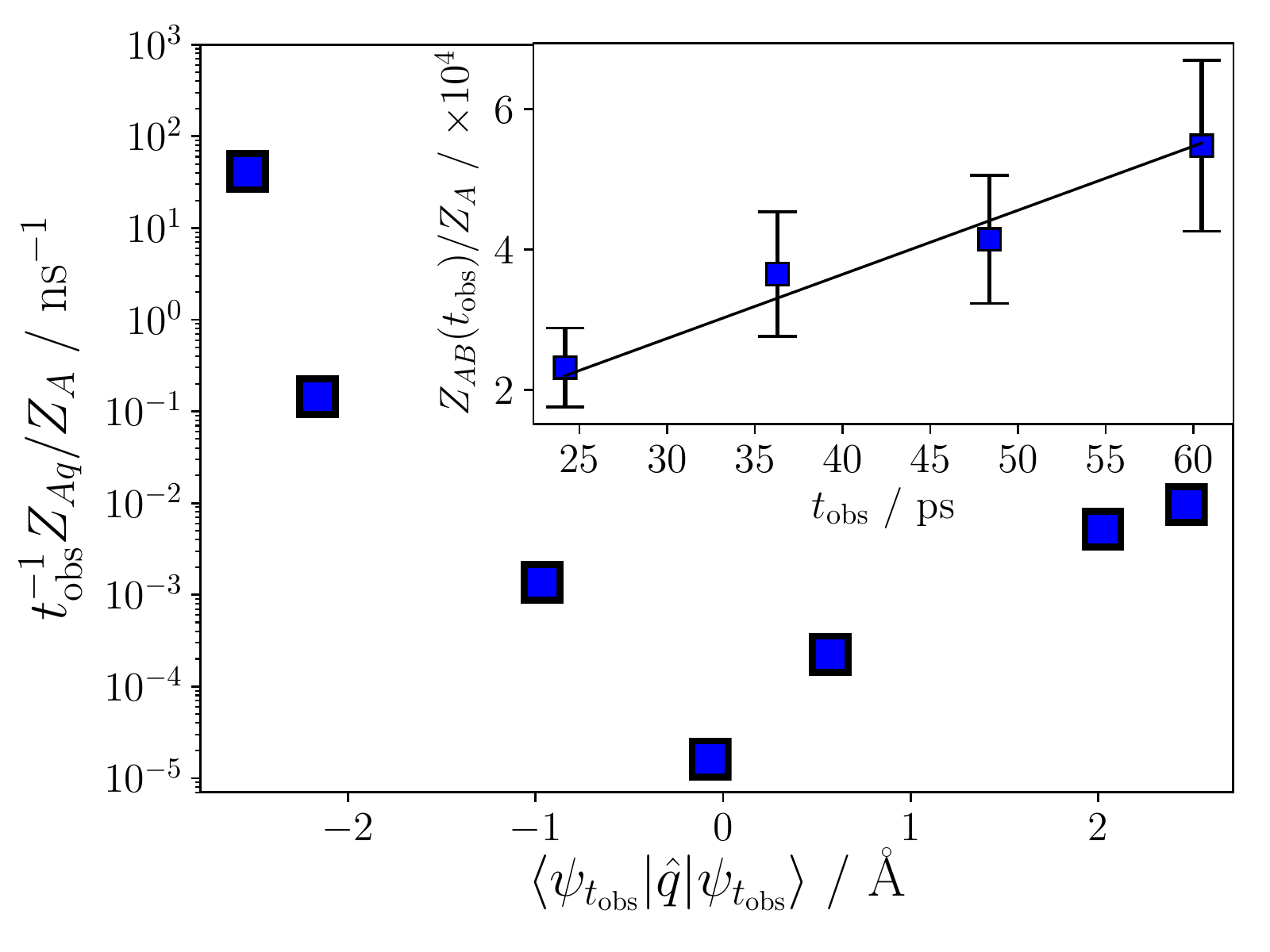}
\caption{Evaluation of the rate using TPS. Ratio of path partition functions computed with $\tobs=24$ ps along the reversible work path. The rightmost point is the rate constant computed from TPS at the observation time. (inset) Ratio of path partition function as a function of $\tobs$. Error bars represent a 95 percent confidence interval computed from block averaging. The black line is a linear fit $k^\mathrm{TPS} \tobs$.}
\label{fig:quartic_rates}
\end{figure}

Specifically, umbrella sampling was performed using overlapping indicator functions, $\op{h}_{\lambda}$, ranging from eigenstates 0-10, with at least one indicator function equal to $\op{h}_{A}=|0\rangle \langle 0| + |2\rangle \langle 2|$ and one equal to $\op{h}_{B}=|1\rangle \langle 1|$. For each window 16,000 trajectories were harvested for every Monte Carlo sweep over an entire trajectory and the expectation value of the position operator $\qexpsym{q}{\tobs}$ corresponding to the eigenstate of the wavefunction at $t=\tobs$ was computed. The statistics of $\langle \psi_{\tobs}  \op{q} | \psi_{\tobs} \rangle$ obtained from this procedure were reweighted using the WHAM procedure~\cite{kumar1992weighted}, which given the discrete outcomes of the observables is a simple optimization routine. This procedure was repeated for a range of values for $\tobs$ from 24 ps to 60 ps. An example of the resulting path partition function ratios for $\tobs = 24$ ps. is shown in Fig. \ref{fig:quartic_rates}. 

These path partition function ratios provide details about the transition rate. First, the ratio divided by $\tobs$ precisely gives the rate of transitions between the reactant state and an intermediate $\lambda$-region provided $\tobs$ is in the linear regime of population transfer. Hence, the rate constant is given in the same thermodynamic language from path ensembles in both the quantum and classical regimes. Finally, the ratio of path partition functions at different values of $\lambda$ offer insight about the mechanism. As $\lambda$ is tuned from eigenstates near the reactant state to the product state the ratio of path free partition functions, as in Fig. \ref{fig:quartic_rates}, decreases indicating a more rare and hence slower rate process, but for eigenstates that are energetically higher than the potential energy barrier, the path partition function ratio is very small, smaller than the ratio for the product state. Hence, states energetically above the potential energy barrier rarely contribute to the predominant transition paths and the typical transitions between the wells are tunneling events.

The resulting rate constant obtained from this umbrella sampling procedure is $k^\mathrm{TPS} =0.010 \pm 0.002$ ns$^{-1}$, which agrees quantitatively with the rate obtained from the population dynamics. Of important note is the short length of trajectories required for computing the rate constant with TPS compared to the population dynamics. Given many accurate quantum dynamics methods have exponential scaling in time, these results suggest that TPS can provide a practical alternative to computing a rate constant to population dynamics. 

For comparison, the rate was also computed using transition state theory (TST), using
\begin{equation}
k^{\text{TST}} = \frac{\omega_0}{2\pi} e^{-\beta \Delta E^{\ddagger}}
\end{equation}
where $\omega_0$ is the frequency of the reactant well, $\Delta E^{\ddagger}$ is the activation energy~\cite{eyring1935activated,wigner1938transition,chandler1978statistical}. The rate obtained by classical TST is 0.0019 ns$^{-1}$, which largely deviates from our result. A temperature-dependent tunneling correction, $\kappa (\beta)$, can also be added, $k = \kappa  (\beta) k^{\text{TST}}$ to account for the tunneling transitions that are predicted by our trajectory analysis. For a parabolic barrier this correction is~\cite{miller1983quantum,thompson1999quantum},
\begin{equation}
\kappa  (\beta) = \frac{\hbar \beta \omega_b / 2}{\sin (\hbar \beta \omega_b /2)},
\end{equation}
here $\omega_b$ is the frequency of the parabolic barrier and corrects the overall rate constant to be 0.011 ns$^{-1}$, which now adds quantitative agreement with the rate obtained from TPS. Such agreement should be expected at low temperature with an approximately parabolic well and barrier as is the case for the quartic potential used here~\cite{thompson1999quantum}. However, in the TPS calculation no assumption about the mechanism was required.

%%%%%%%%%%%%%%%%%%%%%%%%%%%%%%%%%%%%%%%%%%%%%%%%%%%%%%%%%%%%%%%%%%%%%%%%%%%%%
% Conclusion
%%%%%%%%%%%%%%%%%%%%%%%%%%%%%%%%%%%%%%%%%%%%%%%%%%%%%%%%%%%%%%%%%%%%%%%%%%%%%
\section{\label{sec:conclusions}Conclusion}

We have presented a path ensemble formalism useful for the study of quantum dynamics in condensed phases. The formalism enables the computation of conditioned ensembles for typical applications of TPS. To formalize a reactive path ensemble, we required an equation of motion that satisfies detailed balance, the complete positivity of the overall density matrix, and is stochastic. These conditions are satisfied by unravelling a Lindblad master equation into a quantum jump equation. The path ensemble formalism was applied to three systems, for each of which we devised a mapping from the original quantum master equation into a Lindblad form without loss of accuracy. This included developing a stochastic polaronic quantum master, illustrating an ability to invoke weak coupling approximations on transformed Hamiltonians in order to study systems that in the untransformed case were in the strong system-bath coupling regime. The use of conditioned ensembles showed the built-in correlations that can be obtained by sampling biased trajectories. These sorts of correlations could, in principle, be sampled by multi-time correlation functions~\cite{ananth2012flux}, which can be difficult to compute and often require high-level methods due to violations of the quantum regression theorem~\cite{fetherolf2017linear}. Trajectory analysis also enables the identification of transport mechanisms in these systems by sampling the sequence of quantum jumps that occur along trajectories.

We also illustrated how TPS could be used to compute a rate constant. TPS was found to be efficient for sampling rare barrier-crossing trajectories and accurately reproduces the rate constant computed from population dynamics of the reduced density matrix. The necessary trajectory length for quantitative agreement was multiple orders of magnitude less than the reduced density matrix simulation. Other dynamics methods that satisfy properties enabling the path ensemble formalism are applicable~\cite{montoya2015extending,walters2016iterative} and for those methods with a computational complexity that scales with simulation time, TPS may be a key alternative to permit the calculation of rate constants.
While the examples used here are relatively small systems with few degrees of freedom, we expect the utility of the present framework to be clear for large, multidimensional systems. Not only will the calculations be made possible by the reduced scaling of stochastic unraveling, the physical insight gained will become useful in detecting relevant reaction coordinates as the number of potential pathways increase.
\\

\begin{acknowledgments}
This material is based upon work supported by the U.S. Department of Energy, Office of Science, Office of Advanced Scientific Computing Research, Scientific Discovery through Advanced Computing (SciDAC) program under Award Number DE-AC02-05CH11231.  This research used resources of the National Energy Research Scientific Computing Center (NERSC), a U.S. Department of Energy Office of Science User Facility operated under Contract No. DE-AC02-05CH11231.
\end{acknowledgments}

\end{document}